\documentclass[final]{iopart}
\usepackage{bm,amsfonts,amssymb, graphicx}
\usepackage{cite}
\usepackage{epstopdf}

\begin{document}

\title[Harmonic generation in bilayer graphene]{High-order harmonic generation in gapped bilayer graphene}
\author{H K Avetissian, A K Avetissian, A G Ghazaryan,  Kh V Sedrakian, and G F Mkrtchian}

\address{Centre of Strong Fields Physics, Yerevan State University, 1 A. Manukian,
Yerevan 0025, Armenia}

\begin{abstract}
Microscopic nonlinear quantum theory of interaction of coherent
electromagnetic radiation with gapped bilayer graphene is developed. The
Liouville-von Neumann equation for the density matrix is solved numerically
at the multiphoton excitation regime. The developed theory of interaction of
charged carriers with strong driving wave field is valid near the Dirac
points of the Brillouin zone. We consider the harmonic generation process in
the nonadiabatic regime of interaction when the Keldysh parameter is of the
order of unity. On the basis of numerical solutions, we examine the rates of
odd and even high-harmonics at the particle-hole annihilation in the field
of a strong pump wave of arbitrary polarization. Obtained results show that
the gapped bilayer graphene can serve as an effective medium for generation
of even and odd high harmonics in the THz and far infrared domains of
frequencies.
\end{abstract}

\maketitle

%Uncomment for PACS numbers title message

% Keywords required only for MST, PB, PMB, PM, JOA, JOB? 
%\vspace{2pc}
%\noindent{\it Keywords}: Article preparation, IOP journals
% Uncomment for Submitted to journal title message
%\submitto{\JPA}
% Comment out if separate title page not required

\section{Introduction}

High harmonic generation (HHG) is an underlying nonlinear phenomenon at the
interaction of intense electromagnetic radiation with the matter \cite%
{hhg1,Abook}. In the past decades, with the advent of intense laser sources,
HHG has been widely investigated in gaseous medium \cite{hhg2}. These
investigations led to the birth of attosecond physics \cite{att} which makes
possible to directly study the ultrafast atomic and molecular processes on
the subfemtosecond time scale \cite{imig1,imig2,imig3}. The intensity of the
gaseous harmonics is weak because of the low gas density. Therefore it is of
interest to find the ways for HHG in the dense matter. Recently, there has
been successful steps to extend HHG and related processes to bulk crystals 
\cite{sol1,sol2,sol3,sol5,sol6,sol7} and 2D nanostructures, such as graphene
and its derivatives \cite{H2,Mer,H3,H4,H6,H7,H8,H9,H10,H11,H12,H13},
hexagonal boron nitride \cite{BN}, monolayer transition metal
dichalcogenides \cite{TMD}, topological insulator \cite{TI} and buckled 2D
hexagonal nanostructures \cite{Mer2019}. The HHG in solids also can make
possible for studying of the charged carrier dynamics in solids on the
subfemtosecond time scale \cite{corcumsolid}.

Among the mentioned materials graphene \cite{1,2} and few-layer graphene
nanostructures have attracted enormous interest due to their unique physical
properties. Bilayer graphene ($AB$-stacked) \cite{2,1a,1aa} shares many of
the interesting properties of monolayer graphene \cite{9,22b,24b}, but
provides a richer band structure. The interlayer coupling between two
graphene sheets changes the monolayer's Dirac cone, inducing trigonal
warping on the band dispersion and changing the topology of the Fermi
surface. This significantly enhances the rates of HHG \cite{H3} in the THz
region compared to monolayer graphene. Studies of the nonlinear coherent
response in $AB$ -stacked bilayer graphene under the influence of intense
electromagnetic radiation also include modification of quasi-energy
spectrum, the induction of valley polarized currents \cite{26b,27b}, as well
as second- and third- order nonlinear-optical effects \cite{28b,29b,30b,31b}.

The important advantage of bilayer graphene over monolayer one is the
possibility to induce large tunable band gaps under the application of a
symmetry-lowering perpendicular electric field \cite{9,10,16}. Under the
applied perpendicular electric field in plane inversion symmetry is broken
and the topology of bands is also modified. In particular, the bands acquire
Berry curvature \cite{Berry}. Note that with the current technology \cite{16}
one can induce very large gaps $U\simeq 0.28$ $\mathrm{eV}$ in $AB$-stacked
bilayer graphene. The magnitude of such a band gap is sufficient to produce
room-temperature field-effect transistors with a high on-off ratio, which is
not possible in intrinsic graphene materials. The large band gaps can also
make possible effective room temperature HHG in bilayer graphene, which is
suppressed in intrinsic bilayer graphene \cite{H3}. For the gapped
materials, the ionization or electron-hole pair creation is the first step
of HHG . According to Keldysh's seminal papers \cite{Keld1,Keld2}, tunneling
ionization and multiphoton ionization are two main ionization mechanisms
when a gapped sample is exposed to an intense laser field. These regimes are
distinguished by the Keldysh parameter $\gamma _{K}$. In the limit of $%
\gamma _{K}>>1$, the multiphoton ionization dominates in the ionization
process. In the limit of $\gamma _{K}<<1$, the tunneling ionization
dominates. In the so-called nonadiabatic regime $\gamma _{K}\sim 1$, both
multiphoton ionization and tunneling ionization can take place. The most HHG
experiments on atoms fall into the range of tunneling ionization. Note that
in the nonadiabatic regime due to the large ionization probabilities the
intensity of harmonics can be significantly enhanced compared with tunneling
one. From this point of view condensed matter materials and, in particular,
bilayer graphene are preferable due to the tunable band gap with nontrivial
topology.

In the present paper, we develop a nonlinear theory of the gapped bilayer
graphene interaction with coherent electromagnetic radiation. We consider a
multiphoton interaction in the nonadiabatic and nonperturbative regime.
Accordingly, the time evolution of the considered system is found using a
nonperturbative numerical approach, revealing the efficient multiphoton
excitation of a Fermi-Dirac sea in bilayer graphene. We show that there is
intense radiation of harmonics at the pump wave-induced particle/hole
acceleration and annihilation.

The paper is organized as follows. In Sec. II the set of equations for a
single-particle density matrix is formulated and numerically solved in the
multiphoton interaction regime. In Sec. III, we consider the problem of
harmonic generation at the multiphoton excitation of gapped bilayer
graphene. Finally, conclusions are given in Sec. IV.

\section{Multiphoton excitations of Fermi-Dirac sea in gapped bilayer
graphene}

In $AB$-stacked gapped bilayer graphene, the low-energy excitations $%
\left\vert \mathcal{E}_{\sigma }\right\vert <\gamma _{1}\simeq 0.39$ $%
\mathrm{eV}$ in the vicinity of the Dirac points $K_{\zeta }$ (valley
quantum number $\zeta =\pm 1$) can be described by an effective single
particle Hamiltonian \cite{9,22b,24b}:

\begin{equation}
\widehat{H}_{\zeta }=\left( 
\begin{array}{cc}
\frac{U}{2} & g_{\zeta }^{\ast }\left( \mathbf{p}\right) \\ 
g_{\zeta }\left( \mathbf{p}\right) & -\frac{U}{2}%
\end{array}%
\right) ,  \label{1}
\end{equation}%
where 
\begin{equation}
g_{\zeta }\left( \mathbf{p}\right) =-\frac{1}{2m}\left( \zeta \widehat{p}%
_{x}+i\widehat{p}_{y}\right) ^{2}+\mathrm{v}_{3}\left( \zeta \widehat{p}%
_{x}-i\widehat{p}_{y}\right) ,  \label{2}
\end{equation}%
$\mathbf{\hat{p}}=\left\{ \widehat{p}_{x},\widehat{p}_{y}\right\} $\textbf{\ 
}is the electron momentum operator, $m=\gamma _{1}/(2\mathrm{v}_{F}^{2})$ is
the effective mass ($\mathrm{v}_{F}$ is the Fermi velocity in a monolayer
graphene); $\mathrm{v}_{3}=\sqrt{3}a\gamma _{3}/(2\hbar )\approx \mathrm{v}%
_{F}/8$ is the effective velocity related to oblique interlayer hopping $%
\gamma _{3}=0.32$ $\mathrm{eV}$ ($a\approx 0.246$ $\mathrm{nm}$ is the
distance between the nearest $A$ sites). The diagonal elements in Eq. (\ref%
{1}) correspond to opened gap $U$. The first term in Eq. (\ref{2}) gives a
pair of parabolic bands $E=\pm p^{2}/(2m)$, and the second term coming from $%
\gamma _{3}$ causes trigonal warping in the band dispersion. In the
low-energy region the Lifshitz transition (separation of the Fermi surface)
occurs at an energy $\mathcal{E}_{L}=m\mathrm{v}_{3}^{2}/2\simeq 1$ $\mathrm{%
meV}$, and the two touching parabolas are reformed into the four separate
\textquotedblleft pockets\textquotedblright . The spin and the valley
quantum numbers are conserved. There is no degeneracy upon the valley
quantum number $\zeta $, for the issue considered. However, since there are
no intervalley transitions, the valley index $\zeta $ can be considered as a
parameter.

The eigenstates of the effective Hamiltonian (\ref{1}) are the spinors,%
\begin{equation}
\Psi _{\sigma }(\mathbf{r})=\frac{1}{\sqrt{S}}|\sigma ,\mathbf{p}\rangle e^{%
\frac{i}{\hbar }\mathbf{pr}}  \label{3}
\end{equation}%
where%
\begin{equation}
|\sigma ,\mathbf{p}\rangle =\frac{1}{\sqrt{S}}\sqrt{\frac{\mathcal{E}%
_{\sigma }+\frac{U}{2}}{2\mathcal{E}_{\sigma }}}\left( 
\begin{array}{c}
1 \\ 
\frac{1}{\mathcal{E}_{\sigma }+\frac{U}{2}}\Upsilon \left( \mathbf{p}\right)%
\end{array}%
\right) ,  \label{4}
\end{equation}%
corresponding to eigenenergies:%
\begin{equation}
\mathcal{E}_{\sigma }\left( \mathbf{p}\right) =\sigma \sqrt{\frac{U^{2}}{4}%
+\left( \mathrm{v}_{3}p\right) ^{2}-\zeta \frac{\mathrm{v}_{3}p^{3}}{m}\cos
3\vartheta +\left( \frac{p^{2}}{2m}\right) ^{2}}.  \label{5}
\end{equation}%
Here 
\begin{equation}
\Upsilon \left( \mathbf{p}\right) =-\frac{p^{2}}{2m}e^{i2\zeta \vartheta
}+\zeta \mathrm{v}_{3}pe^{-i\zeta \vartheta },  \label{6}
\end{equation}%
$\vartheta =\arctan \left( p_{y}/p_{x}\right) $, $S$ is the quantization
area, and $\sigma $ is the band index: $\sigma =1$ and $\sigma =-1$ for
conduction and valence bands, respectively.

We consider the case when the bilayer graphene interacts with a plane
quasimonochromatic electromagnetic radiation of carrier frequency $\omega $
and slowly varying envelope, and the wave propagates in the perpendicular
direction to the graphene sheets ($XY$) to exclude the effect of the
magnetic field. In general we assume elliptically polarized wave: 
\begin{equation}
\mathbf{E}\left( t\right) =f\left( t\right) E_{0}\left( \widehat{\mathbf{x}}%
\sin \phi \cos \omega t+\widehat{\mathbf{y}}\cos \phi \sin \omega t\right) .
\label{7}
\end{equation}%
The wave envelope is described by the sin-squared envelope function:%
\begin{equation}
f\left( t\right) =\left\{ 
\begin{array}{cc}
\sin ^{2}\left( \pi t/\mathcal{T}_{p}\right) , & 0\leq t\leq \mathcal{T}_{p},
\\ 
0, & t<0,t>\mathcal{T}_{p},%
\end{array}%
\right. ,  \label{8}
\end{equation}%
where $\mathcal{T}_{p}$ characterizes the pulse duration and is taken to be
twenty wave cycles: $\mathcal{T}_{p}=20\mathcal{T}_{0}$, $\phi $ is the pump
wave polarization parameter.

We write the Fermi-Dirac field operator in the form of an expansion in the
free states, given in (\ref{3}), that is, 
\begin{equation}
\widehat{\Psi }(\mathbf{r},t)=\sum\limits_{\mathbf{p,}\sigma }\widehat{a}_{%
\mathbf{p},\sigma }(t)\Psi _{\sigma }(\mathbf{r}),  \label{9}
\end{equation}%
where $\widehat{a}_{\mathbf{p},\sigma }(t)$ ($\widehat{a}_{\mathbf{p},\sigma
}^{+}(t)$) is the annihilation (creation) operator for an electron with
momentum $\mathbf{p}$ which satisfy the usual fermionic anticommutation
rules at equal times. The single-particle Hamiltonian in the presence of a
uniform time-dependent electric field $E(t)$ can be expressed in the form:%
\begin{equation}
\widehat{H}_{s}=\widehat{H}_{\zeta }+\left( 
\begin{array}{cc}
e\mathbf{rE}\left( t\right) & 0 \\ 
0 & e\mathbf{rE}\left( t\right)%
\end{array}%
\right) ,  \label{12}
\end{equation}%
where for the interaction Hamiltonian we have used a length gauge,
describing the interaction by the potential energy \cite{32b,Corkum}. Taking
into account expansion (\ref{9}), the second quantized total Hamiltonian can
be expressed in the form: 
\begin{equation}
\widehat{H}=\sum\limits_{\sigma ,\mathbf{p}}\mathcal{E}_{\sigma }\left( 
\mathbf{p}\right) \widehat{a}_{\sigma \mathbf{p}}^{+}\widehat{a}_{\sigma 
\mathbf{p}}+\widehat{H}_{\mathrm{int}},  \label{13}
\end{equation}%
where the light--matter interaction part is given in terms of the
gauge-independent field $\mathbf{E}\left( t\right) $ as follow:%
\[
\widehat{H}_{\mathrm{int}}=ie\sum\limits_{\mathbf{p,p}^{\prime },\sigma
}\delta _{\mathbf{p}^{\prime }\mathbf{p}}\partial _{\mathbf{p}^{\prime }}%
\mathbf{E}\left( t\right) \widehat{a}_{\mathbf{p},\sigma }^{\dagger }%
\widehat{a}_{\mathbf{p}^{\prime },\sigma ^{\prime }}\ 
\]%
\begin{equation}
\ +\sum\limits_{\mathbf{p},\sigma }\mathbf{E}\left( t\right) \left( \mathbf{D%
}_{\mathrm{t}}\left( \sigma ,\mathbf{p}\right) \widehat{a}_{\mathbf{p}%
,\sigma }^{+}\widehat{a}_{\mathbf{p},-\sigma }+\mathbf{D}_{\mathrm{m}}\left(
\sigma ,\mathbf{p}\right) \widehat{a}_{\mathbf{p},\sigma }^{+}\widehat{a}_{%
\mathbf{p},\sigma }\right) .  \label{14}
\end{equation}%
Here 
\begin{equation}
\mathbf{D}_{\mathrm{t}}\left( \sigma ,\mathbf{p}\right) =\hbar e\langle
\sigma ,\mathbf{p}|i\partial _{\mathbf{p}}|-\sigma ,\mathbf{p}\rangle
\label{15}
\end{equation}%
is the transition dipole moment and 
\begin{equation}
\mathbf{D}_{\mathrm{m}}\left( \sigma ,\mathbf{p}\right) =\hbar e\langle
\sigma ,\mathbf{p}|i\partial _{\mathbf{p}}|\sigma ,\mathbf{p}\rangle
\label{16}
\end{equation}%
is the Berry connection or mean dipole moment.

Multiphoton interaction of a bilayer graphene with a strong radiation field
will be described by the Liouville--von Neumann equation for a
single-particle density matrix%
\begin{equation}
\rho _{\alpha ,\beta }(\mathbf{p},t)=\langle \widehat{a}_{\mathbf{p},\beta
}^{+}\left( t\right) \widehat{a}_{\mathbf{p},\alpha }\left( t\right) \rangle
,  \label{17}
\end{equation}%
where $\widehat{a}_{\mathbf{p},\alpha }\left( t\right) $ obeys the
Heisenberg equation 
\begin{equation}
i\hbar \frac{\partial \widehat{a}_{\mathbf{p},\alpha }\left( t\right) }{%
\partial t}=\left[ \widehat{a}_{\mathbf{p},\alpha }\left( t\right) ,\widehat{%
H}\right] .  \label{18}
\end{equation}%
Note that due to the homogeneity of the problem we only need the $\mathbf{p}$%
-diagonal elements of the density matrix. We will also incorporate
relaxation processes into Liouville--von Neumann equation with inhomogeneous
phenomenological damping term, since homogeneous relaxation processes are
slow compared with inhomogeneous. Thus, taking into account Eqs. (\ref{13})-(%
\ref{18}), the evolutionary equation will be%
\[
i\hbar \frac{\partial \rho _{\alpha ,\beta }(\mathbf{p},t)}{\partial t}%
-i\hbar e\mathbf{E}\left( t\right) \frac{\partial \rho _{\alpha ,\beta }(%
\mathbf{p},t)}{\partial \mathbf{p}}= 
\]%
\[
\left( \mathcal{E}_{\alpha }\left( \mathbf{p}\right) -\mathcal{E}_{\beta
}\left( \mathbf{p}\right) -i\hbar \Gamma \left( 1-\delta _{\alpha \beta
}\right) \right) \rho _{\alpha ,\beta }(\mathbf{p},t) 
\]%
\[
+\mathbf{E}\left( t\right) \left( \mathbf{D}_{\mathrm{m}}\left( \alpha ,%
\mathbf{p}\right) -\mathbf{D}_{\mathrm{m}}\left( \beta ,\mathbf{p}\right)
\right) \rho _{\alpha ,\beta }(\mathbf{p},t) 
\]%
\begin{equation}
+\mathbf{E}\left( t\right) \left[ \mathbf{D}_{\mathrm{t}}\left( \alpha ,%
\mathbf{p}\right) \rho _{-\alpha ,\beta }(\mathbf{p},t)-\mathbf{D}_{\mathrm{t%
}}\left( -\beta ,\mathbf{p}\right) \rho _{\alpha ,-\beta }(\mathbf{p},t)%
\right] .  \label{19}
\end{equation}%
Here $\Gamma $ is the damping rate. In Eq. (\ref{19}) the diagonal elements
represent particle distribution functions for conduction $N_{c}(\mathbf{p}%
,t)=\rho _{1,1}(\mathbf{p},t)$ and valence $N_{\mathrm{v}}(\mathbf{p}%
,t)=\rho _{-1,-1}(\mathbf{p},t)$ bands, and the nondiagonal elements are
interband polarization $\rho _{1,-1}(\mathbf{p},t)=P(\mathbf{p},t)$ and its
complex conjugate $\rho _{-1,1}(\mathbf{p},t)=P^{\ast }(\mathbf{p},t)$.
Thus, we need to solve the closed set of differential equations for these
quantities:%
\[
i\hbar \frac{\partial N_{c}(\mathbf{p},t)}{\partial t}-i\hbar e\mathbf{E}%
\left( t\right) \frac{\partial N_{c}(\mathbf{p},t)}{\partial \mathbf{p}}= 
\]%
\begin{equation}
\mathbf{E}\left( t\right) \mathbf{D}_{\mathrm{t}}\left( \mathbf{p}\right)
P^{\ast }(\mathbf{p},t)-\mathbf{E}\left( t\right) \mathbf{D}_{\mathrm{t}%
}^{\ast }\left( \mathbf{p}\right) P(\mathbf{p},t),  \label{20}
\end{equation}%
\[
i\hbar \frac{\partial N_{\mathrm{v}}(\mathbf{p},t)}{\partial t}-i\hbar e%
\mathbf{E}\left( t\right) \frac{\partial N_{\mathrm{v}}(\mathbf{p},t)}{%
\partial \mathbf{p}}= 
\]%
\begin{equation}
-\mathbf{E}\left( t\right) \mathbf{D}_{\mathrm{t}}\left( \mathbf{p}\right)
P^{\ast }(\mathbf{p},t)+\mathbf{E}\left( t\right) \mathbf{D}_{\mathrm{t}%
}^{\ast }\left( \mathbf{p}\right) P(\mathbf{p},t),  \label{21}
\end{equation}%
\[
i\hbar \frac{\partial P(\mathbf{p},t)}{\partial t}-i\hbar e\mathbf{E}\left(
t\right) \frac{\partial P(\mathbf{p},t)}{\partial \mathbf{p}}= 
\]%
\[
\left[ 2\mathcal{E}_{1}\left( \mathbf{p}\right) +\mathbf{E}\left( t\right) 
\mathbf{D}_{\mathrm{m}}\left( \mathbf{p}\right) -i\hbar \Gamma \right] P(%
\mathbf{p},t) 
\]%
\begin{equation}
+\mathbf{E}\left( t\right) \mathbf{D}_{\mathrm{t}}\left( \mathbf{p}\right) %
\left[ N_{\mathrm{v}}(\mathbf{p},t)-N_{c}(\mathbf{p},t)\right] ,  \label{22}
\end{equation}

\begin{figure}[tbp]
\includegraphics[width=.7\textwidth]{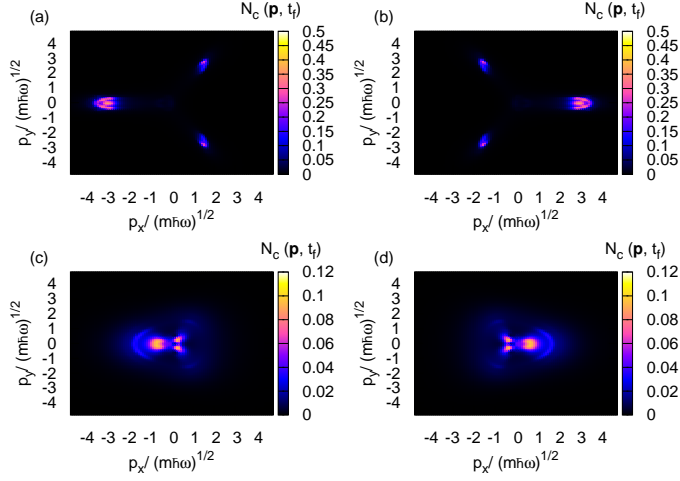}
\caption{(Color online) Particle distribution function $N_{c}(\mathbf{p}%
,t_{f})$\ (in arbitrary units) after the interaction at the instant $%
t_{f}=20T$, as a function of scaled dimensionless momentum components is
shown. The wave is assumed to be linearly polarized along the $y$ axis.
Multiphoton excitation with the trigonal warping effect for the photon
energy $\hbar \protect\omega =E_{L}/1.1$\ $\simeq $\ $0.9$\ $\mathrm{meV}$,
the gap energy $U=4.05$ $\mathrm{meV}$ and the temperature $T/\hbar \protect%
\omega =0.01$ are demonstrated at dimensionless intensity parameter $\protect%
\chi =0.5$ for valleys (a) $\protect\zeta =1$ and (b) $\protect\zeta =-1$.
In (c) and (d) corresponding to valleys $\protect\zeta =1$ and $-1$,
respectively, density plot of the distribution functions are shown for the
photon energy $\hbar \protect\omega $\ $=$\ $50$\ $\mathrm{meV}\simeq
50E_{L} $, energy gap $U=250$ $\mathrm{meV}$, temperature $T/\hbar \protect%
\omega =0.1$, and dimensionless parameter $\protect\chi =1$.}
\end{figure}
\qquad \qquad \qquad

As an initial state we assume an ideal Fermi gas in equilibrium with
vanishing chemical potential and we will solve the set of Eqs. (\ref{20}), (%
\ref{21}), and (\ref{22}) with the initial conditions:%
\begin{equation}
P(\mathbf{p},0)=0;\ N_{c}(\mathbf{p},0)=\frac{1}{1+e^{\mathcal{E}_{1}\left( 
\mathbf{p}\right) /T}};  \label{22a}
\end{equation}%
\begin{equation}
\ N_{\mathrm{v}}(\mathbf{p},0)=1-N_{c}(\mathbf{p},0).  \label{23}
\end{equation}%
Here $T$ is the temperature in energy units.

The components of the transition dipole moments are calculated via Eq. (\ref%
{15}) by spinor wave functions (\ref{4}):%
\[
D_{\mathrm{t}x}\left( \mathbf{p}\right) =-\frac{e\hbar }{2\mathcal{E}%
_{1}\left( \mathbf{p}\right) \sqrt{\mathcal{E}_{1}^{2}\left( \mathbf{p}%
\right) -\frac{U^{2}}{4}}} 
\]%
\[
\mathbf{\times }\left( \left[ \left( \frac{p^{2}}{2m}-m\mathrm{v}%
_{3}^{2}\right) \frac{\zeta p_{y}}{m}+\frac{\mathrm{v}_{3}}{m}p_{x}p_{y}%
\right] \right. 
\]%
\begin{equation}
\left. -i\frac{U}{2\mathcal{E}_{1}}\left\{ \left( \frac{p^{2}}{2m}+m\mathrm{v%
}_{3}^{2}\right) \frac{p_{x}}{m}-\frac{3\zeta \mathrm{v}_{3}}{2m}\left(
p_{x}^{2}-p_{y}^{2}\right) \right\} \right) ,  \label{27}
\end{equation}%
\[
D_{\mathrm{t}y}\left( \mathbf{p}\right) =-\frac{e\hbar }{2\mathcal{E}%
_{1}\left( \mathbf{p}\right) \sqrt{\mathcal{E}_{1}^{2}\left( \mathbf{p}%
\right) -\frac{U^{2}}{4}}} 
\]%
\[
\times \left( \left[ \left( -\frac{p^{2}}{2m}+m\mathrm{v}_{3}^{2}\right) 
\frac{\zeta p_{x}}{m}+\frac{\mathrm{v}_{3}}{2m}\left(
p_{x}^{2}-p_{y}^{2}\right) \right] \right. 
\]%
\begin{equation}
\left. -i\frac{U}{2\mathcal{E}_{1}}\left\{ \left( \frac{p^{2}}{2m}+m\mathrm{v%
}_{3}^{2}\right) \frac{p_{y}}{m}+\frac{3\zeta \mathrm{v}_{3}}{m}%
p_{x}p_{y}\right\} \right) .  \label{28}
\end{equation}%
The total mean dipole moments are%
\[
D_{x\mathrm{m}}\left( \mathbf{p}\right) =D_{x\mathrm{m}}\left( 1,\mathbf{p}%
\right) -D_{x\mathrm{m}}\left( -1,\mathbf{p}\right) =-\frac{e\hbar U}{2%
\mathcal{E}_{1}\left( \mathbf{p}\right) \left( \mathcal{E}_{1}^{2}\left( 
\mathbf{p}\right) -\frac{U^{2}}{4}\right) } 
\]%
\begin{equation}
\mathbf{\times }\left[ \left( \frac{p^{2}}{2m}-m\mathrm{v}_{3}^{2}\right) 
\frac{\zeta p_{y}}{m}+\frac{\mathrm{v}_{3}}{m}p_{x}p_{y}\right] ,  \label{29}
\end{equation}%
\[
D_{y\mathrm{m}}\left( \mathbf{p}\right) =D_{y\mathrm{m}}\left( 1,\mathbf{p}%
\right) -D_{y\mathrm{m}}\left( -1,\mathbf{p}\right) ==-\frac{e\hbar U}{2%
\mathcal{E}_{1}\left( \mathbf{p}\right) \left( \mathcal{E}_{1}^{2}\left( 
\mathbf{p}\right) -\frac{U^{2}}{4}\right) } 
\]%
\begin{equation}
\times \left[ \left( -\frac{p^{2}}{2m}+m\mathrm{v}_{3}^{2}\right) \frac{%
\zeta p_{x}}{m}+\frac{\mathrm{v}_{3}}{2m}\left( p_{x}^{2}-p_{y}^{2}\right) %
\right] .  \label{30}
\end{equation}

\begin{figure}[tbp]
\includegraphics[width=.7\textwidth]{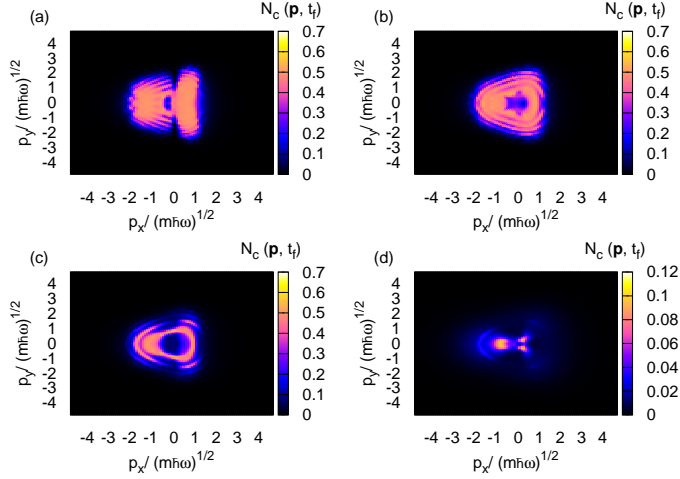}
\caption{ (Color online) Creation of a particle-hole pair in bilayer
graphene at multiphoton resonant excitation. Particle distribution function $%
N_{c}(\mathbf{p},t_{f})$ (in arbitrary units) after the interaction is
displayed at various gap energy: (a) $U=0$, (b) $U=75$ $\mathrm{meV}$, (c) $%
U=125$ $\mathrm{meV}$, and (d) $U=250$ $\mathrm{meV}$. The temperature is
taken to be $T/\hbar \protect\omega =0.1$. The wave is assumed to be
linearly polarized along the $y$ axis with the frequency $\protect\omega =$\ 
$50$\ $\mathrm{meV/\hbar }$ and intensity parameter is $\protect\chi =1$.
The results are for the valley $\protect\zeta =1$. }
\label{222}
\end{figure}

Note that the matrix elements (\ref{27})-(\ref{30}) are actually gauge
dependent. Different choices of the basic function (\ref{4}) (with the phase
factor $e^{i\phi _{\sigma }(\mathbf{p})}$) will not change the energy
spectrum (\ref{5}), but will lead to the different dipole moments. Thus,
inclusion of Berry connection (\ref{29}), (\ref{30}) into dynamics is
mandatory for providing gauge invariance of the final results \cite{Mer2019}.

The set of equations (\ref{20}), (\ref{21}), and (\ref{22}) can not be
solved analytically. For the numerical solution we made a change of
variables and transform the equations with partial derivatives into ordinary
ones. The new variables are $t$ and $\widetilde{\mathbf{p}}=\mathbf{p}-%
\mathbf{p}_{E}$ $\left( t\right) $, where \ 
\begin{equation}
\mathbf{p}_{E}\left( t\right) =-e\int_{0}^{t}\mathbf{E}\left( t^{\prime
}\right) dt^{\prime }  \label{24}
\end{equation}%
is the classical momentum given by the wave field. After these
transformations, the integration of equations (\ref{20}), (\ref{21}), and (%
\ref{22}) is performed on a homogeneous grid of $10^{4}$ ($\widetilde{p}_{x},%
\widetilde{p}_{y}$)-points. For the maximal momentum we take $\widetilde{p}%
_{\max }/\sqrt{m\hbar \omega }=5$. The time integration is performed with
the standard fourth-order Runge-Kutta algorithm. For the relaxation rate we
take $\Gamma =0.5\mathcal{T}^{-1}$. The interaction parameters are chosen as
follow. The wave-particle interaction at the photon energies $\hbar \omega
>E_{L}$\ for intraband transitions can be characterized by the dimensionless
parameter $\chi =eE_{0}/(\omega \sqrt{m\hbar \omega })$, which is the ratio
of the amplitude of the momentum given by the wave field to momentum at
one-photon absorption. Here the intraband transitions are characterized by
the classical momentum given by the wave field $\mathbf{p}_{E}\left(
t\right) $. Besides, due to the gap the interband transitions will be
characterized by the known Keldysh parameter \cite{Keld2}:\textrm{\ }%
\[
\gamma _{K}=\frac{\omega \sqrt{mU}}{eE_{0}}=\frac{1}{\chi }\sqrt{\frac{U}{%
\hbar \omega }}, 
\]%
which governs ionization process in the strong laser fields. For the
considered case the ionization process reduces to the transfer of the
electron from the valence band into the conduction band, in other words, to
the creation of an electron-hole pair. It is obvious that interband
transitions can be neglected when $\gamma _{K}>>1$. The latter means that
wave field can not provide enough energy for the creation of an
electron-hole pair and the generation of harmonics is suppressed. When $%
\gamma _{K}\sim 1$ or $\gamma _{K}<<1$ interband transitions take place. In
the latter case, the transitions correspond to tunneling regime and are
independent on the wave frequency. In the current paper, we will consider
the optimal regime for the generation of harmonics $\gamma _{K}\sim 1$ and $%
\chi \sim 1$.\textrm{\ }Note that the intensity of the wave can be estimated
as\textrm{\ }$I_{\chi }=\chi ^{2}\times 6\times 10^{10}$\textrm{\ }$Wcm^{-2}$%
\textrm{\ }$(\hbar \omega /\mathrm{eV})^{3}$, so the required intensity $%
I_{\chi }$\ for the nonlinear regime$\ $strongly depends\textit{\ }on the
photon energy. 
\begin{figure}[tbp]
\includegraphics[width=.7\textwidth]{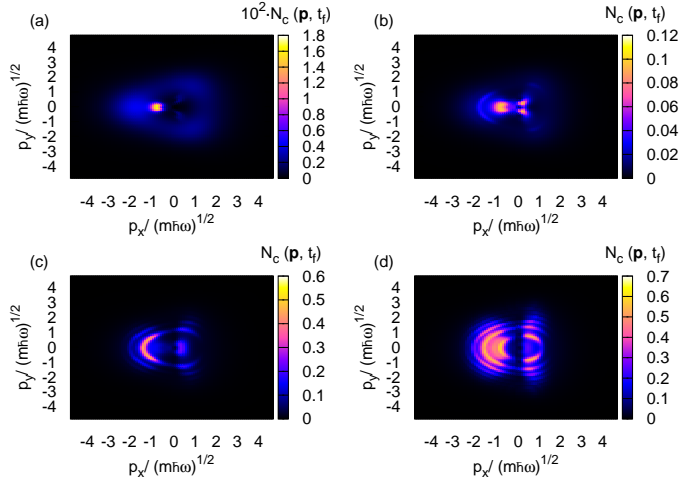}
\caption{(Color online) Creation of a particle-hole pair in bilayer graphene
at multiphoton excitation. Particle distribution function $N_{c}(\mathbf{p}%
,t_{f})$ (in arbitrary units) after the interaction is displayed for various
wave intensities. The temperature is taken to be $T/\hbar \protect\omega %
=0.1 $, and the band gap $U=250$ $\mathrm{meV}$. The wave is assumed to be
linearly polarized along the $y$ axis with the frequency $\protect\omega =$\ 
$50$\ $\mathrm{meV/\hbar }$. The results are for the valley $\protect\zeta %
=1 $: (a)--(d) correspond to dimensionless field parameters $\protect\chi %
=0.5$, $1$, $1.5$, and $2$, respectively. }
\end{figure}

\begin{figure}[tbp]
\includegraphics[width=.7\textwidth]{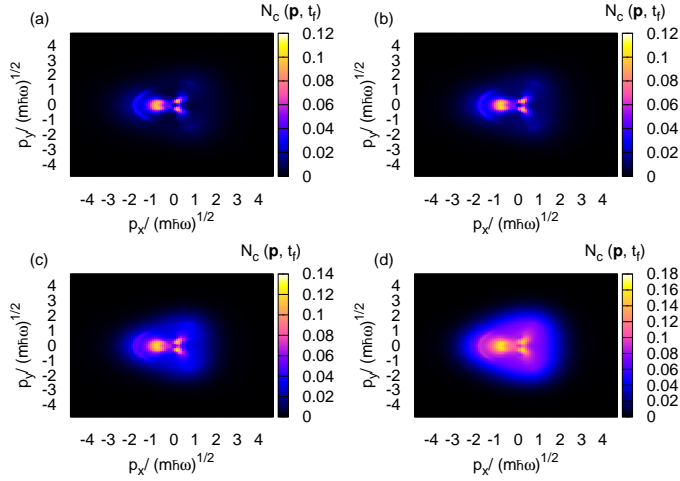}
\caption{ (Color online) Creation of a particle-hole pair in bilayer
graphene at various temperatures and band gap $U=250$ $\mathrm{meV}$.
Particle distribution function $N_{c}(\mathbf{p},t_{f})$ (in arbitrary
units) the results for the valley $\protect\zeta =1$ after the interaction
are displayed for (a) $T/\hbar \protect\omega =0.1$, (b) $T/\hbar \protect%
\omega =0.3$, (c) $T/\hbar \protect\omega =0.5$, and (d) $T/\hbar \protect%
\omega =0.7$. The wave is assumed to be linearly polarized along the y axis
with the frequency $\protect\omega =$\ $50$\ $\mathrm{meV/\hbar }$ at wave
intensity correspond to $\protect\chi =1$. }
\label{1111}
\end{figure}

Photoexcitations of the Fermi-Dirac sea are presented in Figs. 1--4. The
wave is assumed to be linearly polarized along the $y$ axis. Similar
calculations for a wave linearly polarized along the $x$\ axis show
qualitatively the same picture. In Fig.1 density plot of the particle
distribution function $N_{c}(\mathbf{p},t_{f})$\ is shown as a function of
scaled dimensionless momentum components after the interaction. It is
clearly seen the trigonal warping effect describing the deviation of the
excited iso-energy contours from circles. Note that trigonal warping is
crucial for even-order nonlinearity. In Fig. 2 the dependence of the
photoexcitation of the Fermi-Dirac sea on the energy gap is shown. As is
seen with the increasing of $U$ we approach to perturbative regime $\gamma
_{K}>1$ and only weak excitation of Fermi-Dirac sea. In Fig. 3 we show the
photoexcitation depending on the pump wave intensity. For the large values
of $\chi $ when $\gamma _{K}=1.1$ we clearly see multiphoton excitations.
With the increasing wave intensity, the states with absorption of more
photons appear in the Fermi-Dirac sea. The multiphoton excitation of the
Fermi-Dirac sea takes place along the trigonally warped isolines of the
quasienergy spectrum modified by the wave field. Thus, the multiphoton
probabilities of particle-hole pair production will have maximal values for
the iso-energy contours defined by the resonant condition:%
\[
\frac{1}{\mathcal{T}}\int\limits_{0}^{\mathcal{T}}2\mathcal{E}_{1}\left( 
\widetilde{\mathbf{p}}+\mathbf{p}_{E}\left( t\right) ,t\right) dt=n\hbar
\omega ,\ \ n=1,2,3..., 
\]%
These contours are also seen in Fig. 2. The temperature dependence of
excitation of Fermi-Dirac sea is shown in Fig. 4. We see that excited
isolines are slightly smeared out. This effect is small since $U>>$ $T$ and
one can expect that harmonic spectra will be robust against temperature
change in contrast to $U=0$ case where harmonics radiation is suppressed
with the increase of temperature.

\section{Generation of harmonics at the particle-hole multiphoton excitation}

In this section we examine the nonlinear response of bilayer graphene
considering nonadiabatic regime of harmonics generation when the Keldysh
parameter is of the order of unity. For the coherent part of the radiation
spectrum, one needs the mean value of the current density operator,%
\begin{equation}
j_{\zeta }=-2e\left\langle \widehat{\Psi }(\mathbf{r},t)\left\vert \widehat{%
\mathbf{v}}_{\zeta }\right\vert \widehat{\Psi }(\mathbf{r},t)\right\rangle ,
\label{50}
\end{equation}%
where $\widehat{\mathbf{v}}_{\zeta }=\partial \widehat{H}/\partial \widehat{%
\mathbf{p}}$ is the velocity operator and we have taken into account the
spin degeneracy factor $2$. For the effective $2\times 2$ Hamiltonian (\ref%
{1}) the velocity operator in components reads:%
\begin{equation}
\widehat{\mathrm{v}}_{\zeta x}=\zeta \left( 
\begin{array}{cc}
0 & -\frac{1}{m}\left( \zeta \widehat{p}_{x}-i\widehat{p}_{y}\right) +%
\mathrm{v}_{3} \\ 
-\frac{1}{m}\left( \zeta \widehat{p}_{x}+i\widehat{p}_{y}\right) +\mathrm{v}%
_{3} & 0%
\end{array}%
\right) ,  \label{51}
\end{equation}%
\begin{equation}
\widehat{\mathrm{v}}_{\zeta y}=i\left( 
\begin{array}{cc}
0 & \frac{1}{m}\left( \zeta \widehat{p}_{x}-i\widehat{p}_{y}\right) +\mathrm{%
v}_{3} \\ 
-\frac{1}{m}\left( \zeta \widehat{p}_{x}+i\widehat{p}_{y}\right) -\mathrm{v}%
_{3} & 0%
\end{array}%
\right) .  \label{52}
\end{equation}%
Using the Eqs. (\ref{50})--(\ref{52}) and (\ref{17}), the expectation value
of the current for the valley $\zeta $ can be written in the form:%
\[
\mathbf{j}_{\zeta }\left( t\right) =-\frac{2e}{(2\pi \hbar )^{2}}\int d%
\mathbf{p}\left\{ \mathbf{V}\left( \mathbf{p}\right) \left( N_{c}(\mathbf{p}%
,t)-N_{\mathrm{v}}(\mathbf{p},t)\right) \right. 
\]%
\begin{equation}
\left. +2\hbar ^{-1}i\mathcal{E}_{1}\left( \mathbf{p}\right) \left[ \mathbf{D%
}_{\mathrm{t}}\left( \mathbf{p}\right) P^{\ast }(\mathbf{p},t)-\mathbf{D}_{%
\mathrm{t}}^{\ast }\left( \mathbf{p}\right) P(\mathbf{p},t)\right] \right\} ,
\label{53}
\end{equation}%
where 
\begin{equation}
\mathbf{V}\left( \mathbf{p}\right) =\frac{\mathrm{v}_{3}\mathbf{p}-3\zeta 
\frac{\mathrm{v}_{3}p}{2m}\mathbf{p}\cos 3\vartheta +3\zeta \frac{\mathrm{v}%
_{3}p^{3}}{2m}\sin 3\vartheta \frac{\partial \vartheta }{\partial \mathbf{p}}%
+2\frac{\mathbf{p}^{3}}{\left( 2m\right) ^{2}}}{\mathcal{E}_{1}\left( 
\mathbf{p}\right) }.  \label{Vp}
\end{equation}%
is the intraband velocity. As is seen from Eq. (\ref{53}), the surface
current provides two sources for the generation of harmonic radiation. The
first term is the intraband current $\sim \left( N_{c}(\mathbf{p},t)-N_{%
\mathrm{v}}(\mathbf{p},t)\right) $. Intraband high harmonics are generated
as a result of the independent motion of carriers in their respective bands.
The second term in Eq. (\ref{53}) describes high harmonics which are
generated as a result of recombination of accelerated electron-hole pairs.
Since we are in the nonadiabatic regimes, the contribution of both
mechanisms are essential. 
\begin{figure}[tbp]
\includegraphics[width=.45\textwidth]{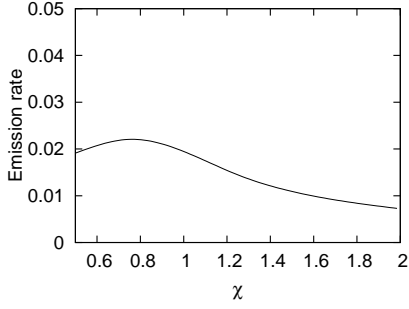}
\caption{Third harmonic scaled emission rate ($G_{3}/\protect\chi ^{3}$) (in
arbitrary units) for bilayer graphene versus $\protect\chi $. The
temperature is taken to be $T/\hbar \protect\omega =0.1$, and the band gap $%
U=250$ $\mathrm{meV}$. The wave is assumed to be linearly polarized along
the $y$ axis with the frequency $\protect\omega =$\ $50$\ $\mathrm{meV/\hbar 
}$. }
\end{figure}
Since there is no degeneracy upon valley quantum number $\zeta $, the total
current is obtained by a summation over $\zeta $: 
\begin{equation}
j_{x}=j_{1,x}+j_{-1,x};  \label{55}
\end{equation}%
\begin{equation}
j_{y}=j_{1,y}+j_{-1,y}.  \label{56}
\end{equation}

From Eq. (\ref{53}) we see that 
\begin{equation}
\frac{j_{x,y}}{j_{0}}=G_{x,y}\left( \omega t,\chi ,\gamma _{K},\frac{%
\mathcal{E}_{L}}{\hbar \omega },\frac{T}{\hbar \omega }\right) ,  \label{57}
\end{equation}%
where 
\begin{equation}
j_{0}=\frac{e\omega }{\pi ^{2}}\sqrt{\frac{m\omega }{\hbar }},  \label{58}
\end{equation}%
and $G_{x}$ and $G_{y}$ are the dimensionless periodic (for monochromatic
wave) functions, which parametrically depend on the interaction parameters $%
\chi $, $\gamma _{K}$, scaled Lifshitz energy, and temperature. Thus, having
solutions of \ Eqs. (\ref{20})-(\ref{22}), and making an integration in Eq. (%
\ref{53}), one can calculate the harmonic radiation spectra with the help of
a Fourier transform of the function $G_{x,y}(t)$. The emission rate of the $%
n $th harmonic is proportional to $n^{2}|j_{n}|^{2}$, where $|j_{n}|^{2}$ = $%
|j_{xn}|^{2}$ + $|j_{yn}|^{2}$, with $j_{xn}$ and $j_{yn}$ being the $n$th
Fourier components of the field-induced total current. To find $j_{n}$, the
fast Fourier transform algorithm has been used. We have used the normalized
current density (\ref{57}) for the plots.

For clarification of the harmonics generation regime we first examine
emission rate of the 3rd harmonic versus pump wave strength $\chi $, which
is shown in Fig. 5. \ As is seen from this figure, up to the field strengths 
$\chi <1$ we almost have power law ($\chi ^{3}$) for the emission rate in
accordance to perturbation theory. For large $\chi $ we have a strong
deviation from power law for the emission rate of 3rd harmonic.

\begin{figure}[tbp]
\includegraphics[width=.45\textwidth]{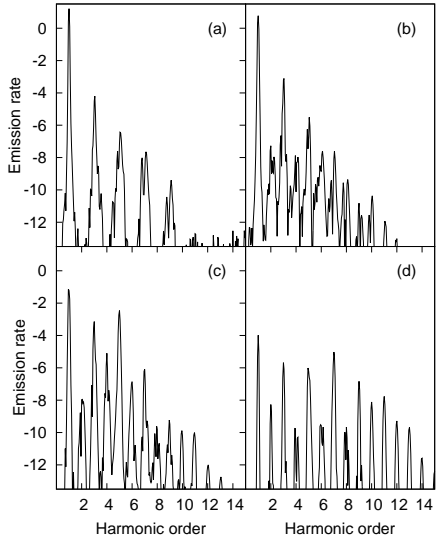}
\caption{ Harmonic emission rate in bilayer graphene at multiphoton
excitation via $log_{10}(n^{2}|G_{n}|^{2})$ (in arbitrary units), as a
function of the photon energy (in units of $\hbar \protect\omega $), is
shown for various gap energies. The temperature is taken to be $T/\hbar 
\protect\omega =0.1$. The wave is assumed to be linearly polarized ($\phi =0$) with the
intensity $\protect\chi =1$ and frequency $\protect\omega =$\ $50$\ $\mathrm{%
meV/\hbar }$. The results are for (a) $U=0$, (b) $U=75$ $\mathrm{meV}$, (c) $%
U=125$ $\mathrm{meV}$, and (d) $U=250$ $\mathrm{meV}$. }
\label{666}
\end{figure}

\begin{figure}[tbp]
\includegraphics[width=.45\textwidth]{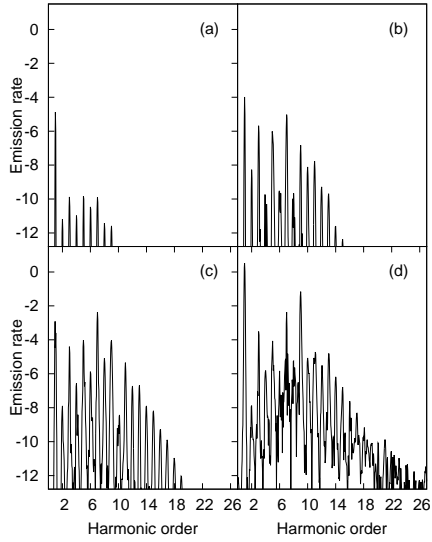}
\caption{High harmonic spectra for bilayer graphene at multiphoton
excitation is shown for various wave intensities in logarithmic scale. The
temperature is taken to be $T/\hbar \protect\omega =0.1$, the gap energy $%
U=250$ $\mathrm{meV}$\textrm{.} The wave is assumed to be linearly polarized ($\phi =0$)
with the frequency $\protect\omega =50$\ $\mathrm{meV/\hbar }$. The results
are for (a) $\protect\chi =0.5$, (b) $\protect\chi =1$, (c) $\protect\chi %
=1.5$, and (d) $\protect\chi =2$. }
\label{777}
\end{figure}

In Fig. 6 the dependence of the harmonic emission rate on the energy gap is
shown. As is seen from this figure, in contrast to intrinsic bilayer
graphene $U=0$ \cite{H3}, when one has in plane inversion symmetry, here
with the increasing of $U$ this symmetry is broken, and as a result, both
even and odd harmonics are emitted. Besides, for the large $U$ due to the
tunneling harmonics the cutoff is increased. In Fig. 7 high harmonic spectra
for bilayer graphene at multiphoton excitation is shown for various wave
intensities. As is seen, with the increasing wave intensity the high-order
harmonics appear in the spectrum. The analysis also shows the linear
dependence of the harmonics number cutoff on the amplitude of a pump
electric field $n_{c}\sim \chi $. The temperature dependence of high
harmonic radiation is clarified in Fig. 7, which shows the robustness of HHG
in gapped graphene against temperature in contrast to intrinsic bilayer
graphene \cite{H3}, where harmonics are suppressed at high temperatures.

\begin{figure}[tbp]
\includegraphics[width=.45\textwidth]{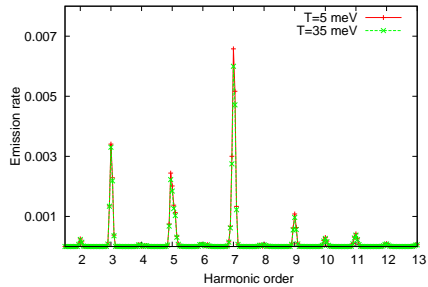}
\caption{ (Color online) High harmonic spectra for bilayer graphene at
multiphoton excitation for a linearly polarized wave ($\phi =0$) is shown at $U=250$ $%
\mathrm{meV}$ for temperatures $T/\hbar \protect\omega =0.7$ and $T/\hbar 
\protect\omega =0.1$. The wave intensity $\protect\chi =1$ and frequency $%
\protect\omega =50$\ $\mathrm{meV/\hbar }$. }
\end{figure}

\begin{figure}[tbp]
\includegraphics[width=.45\textwidth]{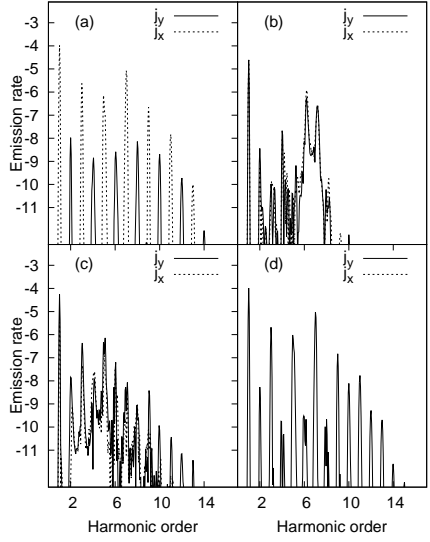}
\caption{High harmonic spectra in logarithmic scale for a elliptically
polarized wave is shown at $U=250$ $\mathrm{meV}$ for temperature $T/\hbar 
\protect\omega =0.1$ at the wave intensity $\protect\chi =1$ and frequency $%
\protect\omega =50$\ $\mathrm{meV/\hbar }$. The results are for (a) $\protect%
\phi =\protect\pi /2$, (b) $\protect\phi =\protect\pi /4$, (c) $\protect\phi %
=\protect\pi /6$, and (d) $\protect\phi =0$, respectively. }
\end{figure}

Finally in Fig. 9 we show the dependence of HHG on the polarization of the
pump wave. The results are for linearly polarized wave along the $x$ ($\phi
=\pi /2$) and $y$ ($\phi =0$) axes, for circular polarization ($\phi =\pi /4$%
) and for elliptic polarization ($\phi =\pi /6$). As is seen, orienting the
linearly polarized pump wave along these axes results\textit{\ }in different
harmonics spectra. This is because we have strongly anisotropic excitation
near the Dirac points. The difference is essential for even-order harmonics.
For elliptic and circular polarizations the rates for the middle harmonics
increase, while high order harmonics are suppressed.

\section{Conclusion}

We have presented the microscopic theory of nonlinear interaction of the
gapped bilayer graphene with a strong coherent radiation field. The energy
gap in considering case is produced by an electric field applied
perpendicular to the bilayer graphene which breaks in plane inversion
symmetry and is modifies the topology of bands. The closed set of
differential equations for the single-particle density matrix is solved
numerically for bilayer graphene in the Dirac cone approximation. For the
pump wave, the THz frequency range has been taken. We have considered
multiphoton excitation of Fermi-Dirac sea towards the high harmonics
generation. It has been shown that the role of the gap in the nonlinear
optical response of bilayer graphene is quite considerable. In particular,
even-order nonlinear processes are present in contrast to intrinsic bilayer
graphene, the cutoff of harmonics increases, and harmonic emission processes
become robust against the temperature increase. The obtained results show
that gapped bilayer graphene can serve as an effective medium for generation
of even and odd high harmonics at room temperatures in the THz and far
infrared domains of frequencies.

This work was supported by the RA MES State Committee of Science and
Belarusian Republican Foundation for Fundamental Research (RB) in the frames
of the joint research projects SCS AB16-19 and BRFFR F17ARM-25, accordingly.

\end{document}